\documentclass[11pt]{article}

\usepackage[utf8]{inputenc}
\usepackage[top=1in,bottom=1in,left=0.75in,right=1in]{geometry}
\usepackage{amssymb}
\usepackage{amsmath}
\usepackage{esvect}
\usepackage{dsfont}
\usepackage{bbold}
\usepackage{slashed}
\usepackage{graphicx}

\newcommand{\be}{\begin{eqnarray}}
\newcommand{\ee}{\end{eqnarray}}

\begin{document}

\title{One-loop Effective Action in Chiral Einstein-Cartan gravity}
\author{Pratik Chattopadhyay \\ \\
\it Department of Physics, Namtech,\\ \it IIT Gandhinagar campus, Gandhinagar 382421, India}

\date{16th October 2023}
\maketitle{}
\begin{abstract} 
In the chiral Einstein-Cartan gravity, a new gauge fixing procedure is implemented recently, leading to a very economical perturbation expansion of the action. Using this formulation and the relevant gauge-fixing, we develop the ghost Lagrangian on an arbitrary Einstein background using the BRST formalism. The novelty is the appearance of a new term quadratic in the tetrad field. We next compute the heat-kernel coefficients and understand the divergences arising in the gravitational one-loop effective action. In our computation the arising heat kernel coefficients depend only on the self-dual part of the Weyl curvature. We make a comparison between our results and what has been obtained for metric GR.
\end{abstract}

\section{Introduction}
One of the major motivations of alternative formulations of General Relativity (GR) is to simplify computations at the perturbative level. In almost all such formulations, one considers more than one independent fields in contrast to just the metric field in usual Einstein gravity. The equivalence of such formulations to metric GR is shown by integrating out the additional fields by using their equations of motion and putting them back to the action. In the tetradic Palatini action \cite{Palatini}, the tetrad and spin connection become independent variables and the metric is understood as a second-order construct. This gives rise to a polynomial action in the absence of the cosmological constant. If one includes the cosmological constant, the action, however, does not remain polynomial. The Einstein-Cartan first-order formulation is the one which keeps the action polynomial when the cosmological constant is nonzero, see \cite{Gronwald:1995em}. A Hamiltonian analysis of first order Einstein-Cartan has been carried out in \cite{Frolov:2009wu}, \cite{Kiriushcheva:2009tg}. The Hamiltonian construct is rather complicated and lead to second class constraints. Further, there has been one-loop computations in this theory in the presence of fermions and the cosmological constant \cite{Shapiro:2014kma}. When fermions are present, it leads to torsion and this enables to enrich the theory by adding a Holst term to the action. 
Inspite of successfully extending usual gravity in the presence of torsion, the first order Einstein-Cartan suffers from its complicated character of the Hamiltonian $(3+1)$ split. Further, at the perturbative level, one has to deal with three propagators which significantly complicate any such perturbative computation. All these difficulties are resolved in the chiral first-order formulation. The original idea of a chiral reformulation of gravity was noted by Plebanski, see \cite{Plebanski:1977zz} and was rediscovered by Capovilla et al \cite{Capovilla:1991qb}. In the chiral Einstein-Cartan as is formulated in \cite{Krasnov:2020bqr}, one considers self-dual projection of the curvature of the spin-connection. One can then construct an action with just the self-dual projection instead of the full curvature. The difference between the two is a total derivative term (Holst term), which does not change the dynamics of the theory. Thus this action is equivalent to the first-order Einstein-Cartan action. Further, one can rewrite the self-dual part of the curvature of the spin connection as the curvature of the self-dual part of the spin connection. This then gives rise to a better perturbation theory where one of the propagators, namely the propagator of the connection with itself, vanishes. Thus the algebraic complexity gets reduced significantly in quantum computations. 
\\~\\
All known reformulations of GR share the feature of non-renormalizability. This is encoded in the negative-dimensional coupling constant, which sits in front of the action. However, as explained in \cite{Groh:2013oaa}, the quantum theory of any such reformulation may not behave in the same way as in the metric GR. Indeed, in the pure connection formulation, the quantum behavior is quite different, with the feature that the sign in front of the $R_{\mu\nu\rho\sigma}^2$ term is opposite to that in the metric GR. 
\\~\\
The purpose of this paper is to study the quantum effects of the chiral Einstein-Cartan action. As we have mentioned, there are different formulations of GR, which are equivalent at the classical level. But if one tries to quantize these theories, one often finds, as in the pure-connection formulation, that the quantum theories are rather different. This makes us ask the question whether all these classically equivalent formulations lead to different quantum theories of gravity. This work is a modest step towards that direction. In particular, we want to understand the one-loop divergences arising in the chiral Einstein-Cartan formalism and compare it with metric GR. To do so, one needs to use the background field method and the relevant heat-kernel technology. One can choose to work with any background, but then one ends up with a very complicated differential operator in the bosonic as well as the ghost sector, which is not minimal (Laplace-type). In an Einstein background, some of these difficulties vanish, and the computation of the heat kernel is simpler for the appearance of just the self-dual part of the Weyl curvature in the differential operator that arises in the bosonic sector. In this calculation, we therefore expand our action around an arbitrary Einstein background, setting the cosmological constant to zero.
\\~\\
When the action is expanded in this background, one gets a novel quadratic term of the tetrad field, which is not present in the flat background case. Another aspect of our calculation is to impose the BRST gauge fixing procedure. Gauge fixing is done in an analogous way to the flat background, and the only difference is that the partial derivatives of \cite{Krasnov:2020bqr} are replaced by background covariant derivatives.
The one-loop computation can be explored using the standard heat-kernel technology as is outlined in \cite{Vassilevich:2003xt}. One first linearizes the action on an Einstein background and then uses the heat kernel to obtain the regularized determinant of the arising operator. As we shall see, the arising operator in the gravitational sector of the action is not of Laplace type. To get a Laplace type operator we square the Dirac operator and then proceed to find the heat-kernel coefficients. In the ghost sector, a different issue arises because in this case the diffeomorphism ghost and the Lorentz ghost are coupled in the Lagrangian. We thus decouple the ghost terms and subsequently get a Laplace-type operator in this sector. All in all, as we shall see, the coefficient of the one-loop divergence arising in this formalism is different than in metric GR and thus it motivates us to perform more non-trivial calculations at more than one-loop and analyze it against what we obtain from metric GR. \\~\\
The paper is organized as follows. In Section 2, we give a brief review of the chiral Einstein-Cartan theory and outline the symmetries in this formalism. In Section 3, we discuss the BRST formalism and explain the gauge-fixing procedure. In Section 4, we detail out the heat kernel computation for both the gravitational and the ghost sectors. We conclude with a discussion of the results obtained. In the appendix, we outline the BRST closure for all the relevant fields.
\section{Review: Chiral Einstein-Cartan theory}
We start with the chiral Einstein Cartan action for gravity, with zero cosmological constant. In units $32\pi G=1$ we have
\be 
\label{ECs}
S[e, \omega]=2i\int \Sigma^{AB}\wedge F_{AB}
\ee 
where $A,B=1,2$ are unprimed 2-component spinor indices and the self-dual 2-forms are 
\be 
\Sigma^{AB}=\frac{1}{2}e^A_{C'}\wedge e^{BC'},
\ee 
where $e^{AA'}$ is the soldering form and the curvature 2-form $F^{AB}$is given by 
\be 
F^{AB}=d\omega^{AB}+\omega^{AC}\wedge \omega^{~B}_C.
\ee 
The object $\omega^{AB}$ is the self-dual part of the spin connection. Locally, it takes values in the spin bundle of symmetric second-rank unprimed spinors. The action in (\ref{ECs}) is obtained by applying the chiral self-dual projection to the first order Einstein-Cartan action in terms of the tetrad $e^{AA'}$ and the full spin connection.
 
\subsection{Symmetries}
The action in (\ref{ECs}) is invariant under two classes of transformations. One is the diffeomorphisms, which is similar to usual Einstein gravity. Another is the local SL(2,C) gauge transformations. When the background is not flat, both these transformations act on the independent fields i.e, the tetrad and the spin connection. As we will see, it is possible to correct the diffeomorphism by a gauge transformation and this results in a simpler set of rules for the transformation of the fields.

\subsubsection{Diffeomorphisms}
Under the infinitesimal coordinate transformation 
\be 
x'^{\mu}=x^{\mu}+\epsilon\xi^{\mu}
\ee 
the tetrad field transforms (at order $\epsilon$) as follows
\be 
\begin{split} 
\delta_{\xi}e^{i}_{\nu}&=\xi^{\mu}\partial_{\mu}e^{i}_{\nu}+e^{i}_{\mu}\partial_{\nu}\xi^{\mu}\\
&=\xi^{\mu}\partial_{\mu}e^{i}_{\nu}-\xi^{\mu}\Gamma^{\lambda}_{\mu\nu}e^i_{\lambda}+e^{i}_{\mu}\partial_{\nu}\xi^{\mu}+e^i_{\mu}\Gamma^{\mu}_{\nu\lambda}\xi^{\lambda}\\
&=\xi^{\mu}\nabla_{\mu}e^{i}_{\nu}+e^{i}_{\mu}\nabla_{\nu}\xi^{\mu}.
\end{split}
\ee 
where in the second line of the above equation, we added and subtracted a term with the Levi-Civita connection, which allows us to write the variation of the tetrad in terms of the spacetime covariant derivative. We can also add and subtract a term with the spin connection. We then have
\be 
\begin{split} 
\delta_{\xi}e^{i}_{\nu}&=\xi^{\mu}\nabla_{\mu}e^{i}_{\nu}+e^{i}_{\mu}\nabla_{\nu}\xi^{\mu}+\xi^{\mu}\omega^{~i}_{\mu~j}e^j_{~\nu}-\xi^{\mu}\omega^{~i}_{\mu~j}e^j_{~\nu}
\\
&=\xi^{\mu}\Big(\nabla_{\mu}e^{i}_{\nu}+\xi^{\mu}\omega^{~i}_{\mu~j}e^j_{~\nu}\Big)+e^{i}_{\mu}\nabla_{\nu}\xi^{\mu}-\xi^{\mu}\omega^{~i}_{\mu~j}e^j_{~\nu}
\\
&=e^{i}_{\mu}\nabla_{\nu}\xi^{\mu}-\xi^{\mu}\omega^{~i}_{\mu~j}e^j_{~\nu}\\
&=e^{i}_{\mu}\nabla_{\nu}\xi^{\mu}-\tilde{\phi}^i_{~j}e^j_{~\nu}
\end{split}
\ee 
where $\xi^{\mu}\omega^{~i}_{\mu~~j}=\tilde{\phi}^{i}_{~j}$. 
Thus, we arrive at a simple transformation rule for the tetrad under diffeomorphisms. The first term is the covariant derivative of the parameter $\xi^{\mu}$ and this is corrected by a gauge transformation. As we will see, this lets us to get simpler rules for the total transformation when we add the local SL(2,C) transformation to it. Let us now write the transformation rule for the connection under diffeomorphisms. The connection is a one-form like the tetrad and thus it will have a similar set of transformations. In particular,
\be 
\begin{split} 
\delta_{\xi}\omega^{~i}_{\mu~j}&=\xi^{\nu}\partial_{\nu}\omega^{~i}_{\mu~j}+\omega^{i}_{\nu~j}\partial_{\mu}\xi^{\nu}.\\
\end{split}
\ee 
We can also write the diffeomorphism transform of the connection in such a way so that it does not contain explicit derivatives of the vector field which generates it. In an analogous way like the tetrad case, we add and subtract pair of terms with a spin connection. We then arrange terms in a way that the vector field is inserted into the curvature of the connection. The remaining term is a total covariant derivative. However, the total covariant derivative term can be matched with the corresponding local SL(2,C) transform of the connection and this results again in a simpler transformation rule. The diffeomorphism of the connection thus reads
\be 
\label{diffconn}
\begin{split} 
\delta_{\xi}\omega^{~i}_{\mu~j}&=\xi^{\nu}F^{i}_{~\nu\mu j}+\nabla^T_{\mu}(\xi^{\nu}\omega^{i}_{\nu~j})
\end{split}.
\ee 
where $\nabla^T_{\mu}$ is the total covariant derivative and the last term in (\ref{diffconn}) is of the form of a gauge transformation, which we can add to the local SL(2,C) part of the transformation of the connection. It is computationally simple to use this version of the transformation. Next, we write the local SL(2,C) transformations of both the tetrad and the connection fields. We consider one chiral half of the Lorentz to act on the fields, the other half being set to zero from the beginning.
\subsubsection{SL(2,C) transformations}
SL(2,C) transformations act on both the tetrad and the connections fields. They are given by 
\be 
\begin{split}
\delta_{\phi}e^{i}_{~\mu}&=\phi^{i}_{~j}e^{j}_{~\mu},\\
\delta_{\phi}\omega^{i}_{\mu~j}&=\nabla^T_{\mu}\phi^{i}_{~j}.
\end{split}
\ee 
Thus the action of the SL(2,C) transformation on the tetrad amounts to a Lorentz rotation given by the intfinitesimal parameter $\phi^i_{~j}$. The action on the connection is given by the covariant derivative on the parameter $\phi^i_{~j}$. 
\subsubsection{Total transformation}
Note that the diffeomorphism of the tetrad is corrected by a gauge transformation, with some parameter $\tilde{\phi}^{ij}$. Then it is possible to absorb this into the local SL(2,C) transformation parameter $\phi^{ij}$ and just set the total gauge transformation parameter to $\phi^{ij}$. Then the full transformation of the tetrad reads
\be 
\delta e^{i}_{~\mu}=e^{i}_{~\nu}\nabla^T_{\mu}\xi^{\nu}-\phi^i_{~j}e^j_{~\mu}.
\ee 
Let us also write the full transformation for the spin connection. We add the diffeomorphism and local Lorentz to get 
\be 
\delta\omega^{i}_{\mu~j}=\nabla^T_{\mu}\phi^{i}_{~j}+\xi^{\nu}F^i_{\nu\mu j}.
\ee 
\subsubsection{Total transformation in spinor notations}
Let us introduce the spinor notations to express the above transformations.
In these notations, the total covariant derivative $\nabla^T_{\mu}$ becomes an object $\nabla^T_{MM'}$ which acts on the spinor version of both Lorentz and spacetime indices, $\omega^{~i}_{\mu~j}$ becomes an object $\omega^{AB}_{~~MM'}$ where $(AB)$ is a pair of symmetrised Lorentz indices and $MM'$ corresponds to the spacetime index $\mu$. The total transformation reads 
\be
\begin{split}
\delta e^{AA'}_{~~MM'}&=e^{AA'}_{~~~NN'}\nabla^T_{MM'}\xi^{NN'}-\phi^{A}_{~~B}e^{BA'}_{~~MM'},\\
\delta\omega^{AB}_{~~~CC'}&=\nabla^T_{CC'}\phi^{AB}+\xi^{MM'}F^{AB}_{0~~MM'CC'}.
\end{split}
\ee 

\section{BRST formalism}
The next step is to quantize the Einstein-Cartan action on an instanton background. The system has two types of gauge symmetries, diffeomorphism and local SL(2,C). It is essential to fix the gauges in order to obtain the physical states of the theory. We will follow the well-developed BRST formalism to gauge fix the theory. Thus, we need to introduce two kinds of ghost fields which enter the BRST transformations. As we will see, the transformation for the local Lorentz (SL(2,C)) ghost will have a term which contains the diffeomorphism ghosts and thus the two ghosts are coupled to generate the BRST complex. This is a nontrivial feature of our formalism. We will gauge fix our theory on a general Einstein background and this makes the BRST closure property more non-trivial to verify. Once we have gauge fixed the theory, we can get the one-loop effective action using the heat-kernel methods. Let us then begin to describe all this.
\subsection{BRST complex}
Consider two pairs of anti-commuting fields $c^{\mu},\bar{c}^{\mu}$ for diffeomorphisms and $b^{ij},\bar{b}^{ij}$ for SL(2,C) transformations. The total BRST transformation operator is  
\be 
s=s_D +s_L,
\ee 
where $s_D$ is the operator corresponding to diffeomorphisms and $s_L$ is for local SL(2,C) transformations. The transformation we use however is most conveniently written in the condensed form using the total transformation operator. Thus the splitting we described above is just to illustrate that it is the total transformation parameter that is nilpotent, i.e $s^2=0$. The individual transformation parameters are not nilpotent and this is the non triviality of the complex in the present case. 
Let us now define the BRST transformations
\be
\label{brstn}
\begin{split}
se^{i}_{~\mu}&=e^{i}_{~\nu}\nabla^T_{\mu}c^{\nu'}-b^{i}_{~j}e^{j}_{~\mu},\\ 
sc^{\mu}&=c^{\nu}\nabla^T_{\nu}c^{\mu},\\
s\bar{c}^{\mu}&=\lambda^{\mu},\\ 
s\lambda^{\mu}&=0,\\
s\omega^{ij}_{~~\mu}&=\nabla^T_{\mu}b^{ij}+c^{\nu}F^{ij}_{~~\nu\mu},\\
sb^{ij}&=-\frac{1}{2}[b,b]^{ij}+\frac{1}{2}c^{\mu}c^{\nu}F^{ij}_{~~\mu\nu},\\
s\bar{b}^{ij}&=\beta^{ij},\\
s\beta^{ij}&=0.
\end{split}
\ee 
As we can see, the transformation for the tetrad has two terms on it, one for diffeomorphism ghost and another for the Lorentz ghost. The relative sign is minus which is purely a convention. In a similar way, the spin connection transformation admits two terms, one of which follows from its Lorentz transformation and another gets added owing to diffeomorphisms. We now verify that the BRST transformation $s$ is nilpotent, i.e $s^2=0$. For the third, fourth, seventh, eighth transformations in (\ref{brstn}), this condition trivially follows because $s^2\lambda^{\mu}=s^2\beta^{ij}=0$ and $s^2\bar{c}^{\mu}=s\lambda^{\mu}=0$, $s^2\bar{b}^{ij}=s\beta^{ij}=0$. For the rest of the transformations, we detail out in Appendix. 
\subsection{Linearised action}
The main objective of the development of BRST complex is to understand the ghost contribution to the one-loop effective action in our chiral formalism. The one loop computation is efficiently done using the heat kernel methods. To employ it one needs to expand the action around an arbitrary background and then compute the regularised determinant of the differential operator which arises. However, in an arbitrary background, the metric need not satisfy the Einstein condition and one cannot use the background field equation to simplify the calculation of the heat kernel. The calculation thus becomes very complicated at this level. Thus, in this paper, we expand the action in (\ref{ECs}) around an arbitrary Einstein background. This simplifies much of the computational complexity which would otherwise had to be dealt with. We then decompose the tetrad and the spin connection into a fixed background and consider small fluctuations around this background.
\be
\label{dec}
e^{AA'}_{~~~MM'}=\tilde{e}^{AA'}_{~~~MM'}+h^{AA'}_{~~~MM'},\ee 
where $\tilde{e}^{AA'}_{~~~MM'}$ is the background tetrad corresponding to an Einstein background, so that the background metric is 
\be 
\tilde{g}_{\mu\nu}=\tilde{e}^{AA'}_{\mu}\tilde{e}_{\nu AA'}.
\ee 
The background metric satisfies Einstein equations. The fluctuations around this metric are then considered and the full metric is decomposed as
\be 
g_{\mu\nu}=\tilde{g}_{\mu\nu}+h_{\mu\nu}. 
\ee 
Similarly, the connection can be split into 
\be 
\label{split}
\omega^{AB}_{T~~CC'}= \omega^{AB}_{0~CC'}+w^{AB}_{~~~CC'}.
\ee 
where $\omega^{AB}_{0~CC'}$ is the background connection.
As a result, the decomposition of the curvature is given by 
\be
\begin{split} 
F_{T}&=F_{0}+\delta F,\\
F^{AB}_{0}&=d\omega^{AB}_0 +\omega_0^{AC}\wedge\omega_{0C}^B,\\
\delta F&=D_{\omega_0}\delta\omega=D_{\omega_0}w,
\end{split}
\ee \\
where in the above, $F_0$ is the background curvature constructed from the background connection, $\delta F$ is the linear order fluctuations around this baackground. The fluctuation can be expressed as a covariant derivative of the connection fluctuation with respect to the background connection. Let us now write the self-dual two form in terms of the splitting form of the tetrad. This will then be used to decompose the action into kinetic and interaction terms in a convenient way.
The self-dual two form can be written as 
\be 
\label{sd}
\begin{split} 
\Sigma^{AB}&=\frac{1}{2}\Big(\tilde{e}^{A}_{~A'}+h^{A}_{~A'}\Big)\wedge\Big(\tilde{e}^{BA'}+h^{BA'}\Big)\\
&=\frac{1}{2}\Big(\tilde{e}^A_{~A'}\wedge \tilde{e}^{BA'}+\tilde{e}^{A}_{~A'}\wedge h^{BA'}+h^{A}_{~A'}\wedge \tilde{e}^{BA'}+h^{A}_{~A'}\wedge h^{BA'}\Big).
\end{split}
\ee
Linearisation of the action in (\ref{ECs}) through the decomposition in (\ref{dec}) is given by 
\be 
\begin{split} 
\label{linear}
S&=i\int \Big(\tilde{e}^A_{~A'}\wedge \tilde{e}^{BA'}+\tilde{e}^{A}_{~A'}\wedge h^{BA'}+h^{A}_{~A'}\wedge \tilde{e}^{BA'}+h^{A}_{~A'}\wedge h^{BA'}\Big)\wedge\Big(d\omega_{0AB}+\omega_{0AC}\wedge \omega^{C}_{0~B}\Big)\\
&+i\int \Big(\tilde{e}^A_{~A'}\wedge \tilde{e}^{BA'}+\tilde{e}^{A}_{~A'}\wedge h^{BA'}+h^{A}_{~A'}\wedge \tilde{e}^{BA'}+h^{A}_{~A'}\wedge h^{BA'}\Big)\wedge\Big(dw_{AB}+\omega_{0AC}\wedge w^{C}_{~B}\\&+w_{AC}\wedge w^{C}_{~B}\Big).
\end{split}
\ee 
where in the above steps we have decomposed the tetrad and the connection fields and written the action in terms of this decomposition. We now expand the wedge product between the tetrad field and the connection and collect all individual terms which can arise. The terms which comprise of just the background fields do not participate in the dynamical equations because they are fixed. Thus we ignore such terms. The rest of the terms have both the background and fluctuations in them. Let us first write the free part of the action. It is given by three terms respectively. 
\be
\begin{split} 
\label{action}
S_{free}&=i\int \tilde{e}^A_{~A'}\wedge \tilde{e}^{BA'}\wedge \Big(w_{AC}\wedge w^{C}_{~B}\Big)+i\int h^{A}_{~A'}\wedge h^{BA'}\wedge\Big(d\omega_{0AB}+\omega_{0AC}\wedge \omega^{C}_{0~B}\Big)\\
&+i\int \Big(\tilde{e}^{A}_{~A'}\wedge h^{BA'}+h^{A}_{~A'}\wedge \tilde{e}^{BA'}\Big)\wedge dw_{AB}.
\end{split} 
\ee 
The first term is quadratic in the connection fluctuations. The appearance of this term makes the tetrad propagator non-zero. The second term is quadratic in the tetrad fluctuations, giving a novel 'mass' term for the gravitons while the third term is a cross term where metric/tetrad fluctuations is wedged with the connection fluctuation. Let us now write the interaction part of the action.
\be 
\begin{split} 
S_{interaction}&=2i\int\tilde{e}^{A}_{~A'}\wedge h^{BA'}\wedge w_{AC}\wedge w^{C}_{~B}\\&+i\int h^{A}_{~A'}\wedge h^{BA'}\wedge\Big(d\omega_{AB}+\omega_{0AC}\wedge \omega^{C}_{~B}+\omega_{AC}\wedge \omega^{C}_{~B}\Big).
\end{split}
\ee 
\\
We next expand the tetrad perturbation and the connection in terms of the background tetrad. This is done to parameterise our tetrad and connection fields in terms of the background and it appears convenient to write the action in terms of the coefficients of the expansion. Since the background is fixed once and for all, we can use it as a basis to expand the fluctuating fields. This is quite similar in spirit to that of the background field method for perturbation theory. The fluctuations of the tetrad and connection then reads
\be 
\begin{split}
\label{expand}
h^{AA'}&=h^{AA'}_{~~~MM'}\tilde{e}^{MM'},\\
\omega^{AB}&=\omega^{AB}_{~~~MM'}\tilde{e}^{MM'},\\
D\omega^{AB}&=\tilde{\nabla}_{MM'}\omega^{AB}_{~~~NN'}\tilde{e}^{MM'}\wedge \tilde{e}^{NN'}.
\end{split}
\ee 
where $\tilde{\nabla}_{MM'}$ is the background covariant derivative. Using the above expansion, we write the kinetic part of the free action as follows 
\be 
\begin{split}
\label{kfree}
S_{kinetic}&=i\int\Big(\tilde{e}^{A}_{~A'}\wedge h^{BA'}+h^{A}_{~A'}\wedge \tilde{e}^{BA'}\Big)\wedge d\omega_{AB}\\
&=i\int(\tilde{e}^{A}_{A'}\wedge \tilde{e}^{MM'}\wedge \tilde{e}^{KK'}\wedge \tilde{e}^{JJ'})h^{BA'}_{~~~MM'}\tilde{\nabla}_{KK'}\omega_{ABJJ'}
\\
&+i\int(\tilde{e}^{MM'}\wedge \tilde{e}^{BA'}\wedge \tilde{e}^{KK'}\wedge \tilde{e}^{JJ'})h^A_{~MM'A'}\tilde{\nabla}_{KK'}\omega_{ABJJ'}.
\end{split}
\ee 
In the above action, we have replaced the derivative on the spin connection by the appropriate background covariant derivative, which acts on the coefficient of the expansion of the connection on the basis of the background tetrads. This then couples with the coefficient of the fluctuation of the tetrad field, thus giving the resultant kinetic terms of the form $hd\omega$ with the background covariant derivatives everywhere in place of the ordinary derivatives.
Next we identify the wedge product of four copies of tetrad as the oriented volume form
\be 
\label{wedge}
\tilde{e}^{AA'}\wedge \tilde{e}^{BB'}\wedge \tilde{e}^{CC'}\wedge \tilde{e}^{DD'}=\epsilon^{AA'BB'CC'DD'}v.
\ee 
and use the spinor representation of the totally anti-symmetric tensor. 
\be 
\label{anti}
\epsilon^{AA'BB'CC'DD'}=i\Big(\epsilon^{AD}\epsilon^{BC}\epsilon^{A'C'}\epsilon^{B'D'}-\epsilon^{AC}\epsilon^{BD}\epsilon^{A'D'}\epsilon^{B'C'}\Big).
\ee 
The validity of the spinor representation can be easily checked. For instance, if we swap any pair of indices, say $AA'\leftrightarrow BB'$, the right hand side picks up an overall minus sign, because the first term of the right side goes to the second under the swap. In a similar way, for all such pairs, an overall sign factor comes up. This shows the totally anti-symmetric nature of the tensor $\epsilon^{AA'BB'CC'DD'}$. Let us then use it and write the kinetic part in (\ref{kfree}) as
\be 
\begin{split} 
\label{kinetic1} 
S_{kinetic}&=\int d^4x\Big[ h^{BJ'JK'}\tilde{\nabla}_{KK'}\omega^K_{~BJJ'}- h^{BK'KJ'}\tilde{\nabla}_{KK'}\omega^J_{~BJJ'}\Big]\\
&+\int d^4x\Big[h^{AKJ'K'}\tilde{\nabla}_{KK'}\omega_{A~JJ'}^{~J}-h^{AJK'J'}\tilde{\nabla}_{KK'}\omega^{~K}_{~AJJ'}\Big]\\
&=-2\int d^4x \Big[ h^{BJ'JK'}\Big(\tilde{\nabla}_{JJ'}\omega^{K}_{~BKK'}-\tilde{\nabla}_{KK'}\omega^K_{~BJJ'}\Big)\Big].
\end{split} 
\ee 
We now decompose the tetrad perturbation into its irreducible components. This is reminiscent of the decomposition of a tensorial quantity, for instance the Riemann tensor, in which we treat the tetrad field as an object with four spinor indices. The decomposition reads
\be 
\label{tetrad dec}
h^{AA'MM'}=h^{(AM)(A'M')}+h^{(AM)}\epsilon^{A'M'}+h^{(A'M')}\epsilon^{AM}+\epsilon^{AM}\epsilon^{A'M'}h.
\ee 
The first object on the right hand side above is a tetrad field with its primed and unprimed indices symmetrised. The object with two spinor indices in the second and third terms are the part of the tetrad perturbation is a symmetric field which only propagate off-shell.
It is then convenient to combine the second and fourth terms and define a new field $h^{AM}=h^{(AM)}+h\epsilon^{AM}$. This field is thus no more symmetric in its pair of unprimed indices. This results in 
\be 
h^{AA'MM'}=h^{(AM)(A'M')}+h^{AM}\epsilon^{A'M'}+h^{(A'M')}\epsilon^{AM}.
\ee 
We get rid of the $h^{A'M'}$ part of the perturbation by setting one chiral half of the Lorentz gauge to zero. This can be done because this part of the perturbation does not appear in the free Lagrangian. Thus it simplifies the computation to some extent. We put the decomposition 
\be 
h^{AA'MM'}=h^{(AM)(A'M')}+h^{AM}\epsilon^{A'M'}
\ee 
in (\ref{kinetic1}) to get for the kinetic part of the Lagrangian 
\be 
\begin{split}
\label{Lk}
\mathcal{L}_{kinetic}&=2h^{JKJ'K'}\Big(\tilde{\nabla}_{KJ'}\omega^M_{~JMK'}-\tilde{\nabla}_{MK'}\omega_{J~KJ'}^{~M}\Big)\\&+2h^{JK}\Big(\tilde{\nabla}_{KJ'}\omega^{M~~J'}_{~JM}-\tilde{\nabla}_{MK'}\omega_{J~K}^{~M~K'}\Big).
\end{split}
\ee 
The kinetic term above can be written in a compact way. It is possible to combine the parts of the tetrad perturbation into a single object and the covariant derivative can then act on an appropriate combination of connection perturbations. This is helpful because it gives rise to a familiar kinetic term of the form $h\nabla\omega$ in a general background which is the one suitable for a chiral first order formulation. Thus, re-written in this way, we get  
\be 
\label{compactke}
\begin{split}
\mathcal{L}_{kinetic}&=-2\Big(h^{JKJ'K'}-h^{(JK)}\epsilon^{J'K'}\Big)\tilde{\nabla}_{MJ'}\Big(\omega_{J~KK'}^{~M}+\epsilon^{M}_{~~K}\omega^N_{~JNK'}\Big)\\&+4h^{JK}\tilde{\nabla}_{KK'}\omega^{M~~K'}_{~JM}.
\end{split}
\ee 
We now redefine some fields as follows 
\be 
\begin{split} 
\omega^{JJ'}&=\omega^{MJ~~J'}_{~~~M},\\
\Omega^{JKMJ'}&=\omega^{JMKJ'}+\epsilon^{M}_{~~K}\omega^{JJ'}.
\end{split}
\ee 
Thus the Lagrangian is written in terms of the new variables 
\be 
\label{compactke}
\begin{split}
\mathcal{L}_{kinetic}&=-2\Big[h^{JKJ'K'}-h^{(JK)}\epsilon^{J'K'}\Big]\tilde{\nabla}_{MJ'}\Omega^{~M}_{J~KK'}+4h^{JK}\tilde{\nabla}_{KJ'}\omega^{~J'}_{J}.
\end{split}
\ee 

As we see, there are two terms in the kinetic part above. However, the kinetic term is degenerate. This is because the connection field $\Omega^{JKMJ'}$ has twelve independent components, while the other field $\omega^{JJ'}$ has four components. The tetrad perturbation $h^{JKJ'K'}$ has nine components, which is conjugate to $\Omega^{JKMJ'}$ and therefore there is a mismatch in the number of components, which makes it degenerate. To remove the degeneracy we need to completely fix the gauge. We want to add a specific gauge fixing and ghost term. This will be done by introducing a gauge fixing fermion in the BRST formalism. Before doing that, let us write down the potential term for the tetrad.
\be 
\mathcal{L}_{tetrad-potential}=h^{A}_{~A'}\wedge h^{BA'}\wedge F_{0AB}
\ee 
For brevity, let us also write the potential terms for the connection. Following \cite{Krasnov:2020bqr}, we have 
\be 
\begin{split}
\mathcal{L}_{connection-potential}&=-\omega^{JKMJ'}\omega_{JKMJ'}+2\omega^{JK~J'}_{~~K}\omega^{M}_{~JMJ'}\nonumber\\&=-\Omega^{JKMJ'}\Omega_{JKMJ'}+2\omega^{JJ'}\omega_{JJ'}
\end{split}
\ee 
Let us now introduce the linearised BRST which we subsequently use to fix our gauges.
\subsection{Linearised BRST}
The main purpose of this section is to disentangle the background from the perturbations while defining the BRST transformations. This will allow us to systematically use the transformations for the perturbations themselves. Let us note how the perturbations of the metric and the connection transform under diffeomorphisms and local SL(2,C) transformations. We begin with the transformation for the general tetrad. It admits the following decomposing into background and quantum fluctuations
\be 
e^{AA'}_{~~~BB'}=\tilde{e}^{AA'}_{~~~BB'}+h^{AA'}_{~~~BB'}.
\ee 
where the perturbation $h^{AA'}_{~~~BB'}$ is of order $\epsilon$ and is very small compared to the background. 
Then, we get for the total transformation of the tetrad perturbation
\be 
\begin{split}
\label{trans}
\delta_{\xi}h^{AA'}_{~~~BB'}&=\tilde{e}^{AA'}_{~~~NN'}\tilde{\nabla}_{BB'}\xi^{NN'}-\phi^A_{~C}\tilde{e}^{CA'}_{~~BB'}.
\end{split}
\ee
where $\tilde{\nabla}_{BB'}$ is the background total covariant derivative.
For the connection field, we similarly decompose it into the background and perturbations
\be 
\label{pert}
\omega^{AB}_{T~~CC'}= \omega^{AB}_{0~CC'}+w^{AB}_{~~~CC'}.
\ee 
where $\omega^{AB}_{0~CC'}$ is the background and $w^{AB}_{~~~CC'}$ is the perturbation of order $\epsilon$.
We get for the transformation of the perturbation, under diffeomorphisms,
\be 
\label{omegavar}
\delta_{\xi}w^{AB}_{~~CC'}&=\xi^{NN'}F^{AB}_{0~~NN'CC'}+\tilde{\nabla}_{CC'}\phi^{AB}.
\ee
where $F^{AB}_{0~~NN'CC'}$ is the background curvature.
With these linearised versions of the diffeomorphisms and gauge transformations, we are now ready to write the BRST transformations of these fields. As usual, we have two pairs of ghost fields, each for diffeomorphisms and local Lorentz transformations. Our linearised BRST is then given by
\be
\label{brstlin}
\begin{split}
sh^{AA'}_{~~MM'}&=\tilde{\nabla}_{MM'}c^{AA'}-b^{A}_{~~B}\tilde{e}^{BA'}_{~~MM'},\\
sc^{MM'}&=c^{LL'}\nabla_{LL'}c^{MM'},\\
s\bar{c}^{MM'}&=\lambda^{MM'},\\ 
s\lambda^{LL'}&=0,\\
sw^{AB}_{~~~CC'}&=c^{MM'}F^{AB}_{0~~MM'CC'}+\tilde{\nabla}_{CC'}b^{AB},\\
sb^{AB}&=-\frac{1}{2}[b,b]^{AB}+\frac{1}{2}c^{MM'}c^{NN'}F^{AB}_{0~~MM'NN'},\\
s\bar{b}^{AB}&=\beta^{AB},\\
s\beta^{A'B'}&=0.
\end{split}
\ee 
\subsection{Gauge fixing}
We follow the gauge fixing procedure which is already described in \cite{Krasnov:2020bqr}. The main difference is that we are dealing with a general Einstein background as opposed to the flat background in \cite{Krasnov:2020bqr}. Thus, instead of linear gauges, there will be non-linearities present in our formalism. We also adopt the BRST gauge fixing procedure, in which we write the gauge fixing fermion and take BRST variations to produce the gauge fixing and ghost terms. This will help us to generate the ghost Lagrangian in a general Einstein background. Let us then begin to fix the gauges. One chiral half of the Lorentz was fixed by making the $h^{A'B'}$ part of the perturbation vanish. The other chiral half of the Lorentz can be fixed by imposing the non-linear version of the Lorentz gauge fixing condition. Consider then the following gauge fixing fermion 
\be 
\label{Lgf}
\begin{split}
\psi_{Lorentz}&=\bar{b}^{JK}\tilde{\nabla}_{MM'}\Omega^{~~~MM'}_{JK}.
\end{split}
\ee 
Using (\ref{brstlin}), the BRST variation of the gauge fixing fermion gives
\be
\begin{split}
s\psi_{Lorentz}&=s\bar{b}^{JK}\tilde{\nabla}_{MM'}\Omega^{~~~MM'}_{JK}-\bar{b}^{JK}\tilde{\nabla}_{MM'}s\Omega^{~~~MM'}_{JK}\\
&=\beta^{JK}\tilde{\nabla}_{MM'}\Omega^{~~~MM'}_{JK}-\bar{b}^{JK}\tilde{\nabla}_{MM'}\tilde{\nabla}^{MM'}b_{JK}\\&-\bar{b}^{JK}\tilde{\nabla}_{MM'}\Big(c^{CC'}F^{MM'}_{0~~JKCC'}\Big).
\end{split}
\ee 
The variation of the Lorentz gauge fixing fermion results in a Lorentz gauge fixing term of the form
\be 
\mathcal{L}_{Lorentz~g.f}=-2\beta^{JK}\tilde{\nabla}_{MM'}\Omega_{JK}^{~~~MM'}.
\ee 
The above term can be written as $-2\beta^{JK}\epsilon^{J'K'}\nabla_{MJ'}\Omega_{J~KK'}^{~M}$, which when added to the Lagrangian in (\ref{compactke}) along with the ghost terms results in
\be 
\label{lorentzgf}
\begin{split}
   \mathcal{L}&=-2\Big[h^{JKJ'K'}+(\beta^{JK}-h^{(JK)})\epsilon^{J'K'}\Big]\tilde{\nabla}_{MJ'}\Omega^{~M}_{J~KK'}+4h^{JK}\tilde{\nabla}_{KJ'}\omega^{~J'}_{J} \\&-\bar{b}^{JK}\tilde{\nabla}_{MM'}\tilde{\nabla}^{MM'}b_{JK}-\bar{b}^{JK}\tilde{\nabla}_{MM'}\Big(c^{CC'}F^{MM'}_{0~~JKCC'}\Big).
\end{split}
\ee
The remaining diffeomorphism gauge can be fixed by implementing a variant of the de-Donder gauge fixing condition as follows. First we introduce a new name for the combination of the terms in the bracketed expression of the first line in (\ref{lorentzgf})
\be 
\label{redf}
H^{JKJ'K'}:=h^{JKJ'K'}+(\beta^{JK}-h^{(JK)})\epsilon^{J'K'}.
\ee 
Note that since the field $\beta^{JK}$ is independent, this results in the two fields $H^{JKJ'K'}$ and $h^{JK}$ being independent of each other. This then leads to decoupling of the kinetic part of the Lagrangian into two sectors, $(H,\Omega)$ and $(h,\omega)$. We now construct the diffeomorphism gauge-fixing fermion to be 
\be 
\label{gff}
\psi_{diffeo}=2\bar{c}^{JJ'}\Big[\tilde{\nabla}_{KK'}\Big(h^{~K'K}_{J~~~J'}+h^{K~~~K'}_{~J'J}\Big)-2\tilde{\nabla}_{KJ'}\beta^{~K}_J-2\beta_{JJ'}\Big].
\ee 
After using the redefinition in (\ref{redf}) we take the BRST variation of it. We have 
\be 
\begin{split} 
\label{brstdiffeo}
s\psi_{diffeo}&=4\beta^{JJ'}\Big[\tilde{\nabla}^{KK'}H_{JKJ'K'}-\tilde{\nabla}^K_{~J'}H^{~~M'}_{JK~M'}+\tilde{\nabla}^K_{~J'}h_{JK}-\beta_{JJ'}\Big]\\
&-4\bar{c}^{JJ'}\Big[\tilde{\nabla}^{KK'}\Big(\tilde{\nabla}_{KK'}c_{JJ'}-b_{JK}\epsilon_{J'K'}\Big)\Big].
\end{split}
\ee 
The variation produces a diffeomorphism gauge fixing term and a ghost term. The gauge fixing term is an analogue of the variant of de-Donder gauge in curved spacetime. The ghost term consists of a kinetic part for the diffeomorphism ghost and a mixed part where the diffeomorphism and Lorentz ghosts couple.
Then, the total gauge fixing term becomes 
\be
\begin{split} 
\label{totalgf}
\mathcal{L}_{g.f}&=4\beta^{JJ'}\Big[\tilde{\nabla}^{KK'}H_{JKJ'K'}-\tilde{\nabla}^{K}_{~J'}H^{~~M'}_{JK~M'}+\tilde{\nabla}^{K}_{~J'}h_{JK}-\beta_{JJ'}\Big]\\&-2\beta^{JK}\tilde{\nabla}_{MM'}\Omega_{JK}^{~~~MM'}.
\end{split} 
\ee 
and the ghost Lagrangian is given by 
\be 
\label{ghost}
\begin{split} 
\mathcal{L}_{ghost}&=-4\bar{c}^{JJ'}\tilde{\nabla}^{KK'}\tilde{\nabla}_{KK'}c_{JJ'}+4\bar{c}^{JK'}\tilde{\nabla}_{KK'}b^{K}_{~J}-\bar{b}^{JK}\tilde{\nabla}_{MM'}\tilde{\nabla}^{MM'}b_{JK}\\&-\bar{b}^{JK}\tilde{\nabla}_{MM'}\Big(c^{CC'}F^{MM'}_{0~~JKCC'}\Big).
\end{split}
\ee 
The Lagrangian above has terms with Lorentz and diffeomorphism ghosts mixed. To compute the ghost contribution to the one-loop effective action, we need to diagonalize and rid of the mixed terms. First, we re-scale the fields 
\be 
\begin{split} 
\label{rescale} 
c^{AA'}&\rightarrow 2c^{AA'},~~\bar{c}^{AA'}\rightarrow 2\bar{c}^{AA'}.
\end{split} 
\ee 
and rewrite the Lagrangian 
\be 
\label{ghost22}
\begin{split} 
\mathcal{L}_{ghost}&=-\bar{c}^{JJ'}\tilde{\nabla}^{KK'}\tilde{\nabla}_{KK'}c_{JJ'}+2\bar{c}^{JK'}\tilde{\nabla}_{KK'}b^{K}_{~J}-\bar{b}^{JK}\tilde{\nabla}_{MM'}\tilde{\nabla}^{MM'}b_{JK}\\&-\frac{1}{2}\bar{b}^{JK}\tilde{\nabla}_{MM'}\Big(c^{CC'}F^{MM'}_{0~~JKCC'}\Big).
\end{split}
\ee 
We now use the background field equation and write 
\be
\begin{split}
\label{psi}
F_{0~~MM'NN'}^{AB}&= \psi^{AB}_{~~CD}\Sigma^{CD}_{0~~MM'NN'}
=\psi^{AB}_{~~MN}\epsilon_{M'N'}.
\end{split}
\ee 
where $\psi^{ABCD}$ is the self-dual part of the Weyl tensor and the self-dual two-form $\Sigma_0$ has the particular representation in terms of the epsilons, given in the second line of the above equation. When we contract all the indices of $F_{0}^{AB}$, we end up with contracting all the indices of $\psi^{ABCD}$, which implies vanishing of the self-dual part of the Weyl tensor. 
\be 
\begin{split} 
F_0&=F_{0~~MM'NN'}^{AB}\epsilon^{N}_{~A}\epsilon^{M}_{B}\epsilon^{M'N'}\\
&=\psi^{AB}_{~~MN}\epsilon_{M'N'}\epsilon^{N}_{~A}\epsilon^{M}_{~B}\epsilon^{M'N'}=0.
\end{split}
\ee
where we used the fact that the self-dual part of the Weyl tensor vanishes upon index contraction.
The Lagrangian in (\ref{ghost22}) can now be written by replacing $F_0$ with the self-dual part of the Weyl tensor. We have  
\be 
\label{ghost3!}
\begin{split} 
\mathcal{L}_{ghost}&=-\bar{c}^{JJ'}\tilde{\nabla}^{KK'}\tilde{\nabla}_{KK'}c_{JJ'}+2\bar{c}^{JK'}\tilde{\nabla}_{KK'}b^{K}_{~J}-\bar{b}^{JK}\tilde{\nabla}_{MM'}\tilde{\nabla}^{MM'}b_{JK}\\&-\frac{1}{2}\psi^{M}_{~~CJK}\bar{b}^{JK}\tilde{\nabla}_{MM'}c^{CM'}.
\end{split}
\ee 

\section{Heat Kernel computation}
Let us consider a Laplace-type operator of the form $\Delta=\mathcal{D}^{\mu}\mathcal{D}_{\mu}+E$. The operator acts on some vector bundle, $E$ is an endomorphism on the fibre and $\mathcal{D}_{\mu}$ is an appropriate covariant derivative. The interesting object to study is the determinant $\textrm{det}(\Delta)$. We use the identity log det($\Delta$)=Tr log($\Delta$) and then rewrite the logarithm of the determinant in terms of an integral  
\be 
\label{logdet}
\textrm{log det}(\Delta)=-\int^{\infty}_0\frac{dt}{t}\textrm{Tr}(e^{-t\Delta}).
\ee 
Then we have the well-known expansion of the trace under the integral, in powers of the auxiliary variable $t$
\be 
\label{trace}
\textrm{Tr}(e^{-t\Delta})=\int d^4x \sqrt{g}\frac{1}{(4\pi t)^2}\sum_{n=0}^{\infty} t^n a^{\mathcal{R}}_n(E).
\ee 
Our task is to compute the heat kernel coefficients $a^R_n(E)$. In the chiral Einstein-Cartan theory around some Einstein background, we consider a manifold $M$ without a boundary, over which we have a vector bundle $V$. The Laplace operator $\mathcal{D}_{\mu}\mathcal{D}^{\mu}$ acts on $V$. Then there exists an expansion of the heat kernel coefficients such that all the odd indexed coefficients vanish and the even indexed coefficients are given by geometric invariants. In particular, the divergent UV behavior is controlled by the coefficient $a^R_2(E)$, which is given by
\be
\label{coeff}
\begin{split} 
a^\mathcal{R}_2(E)&=Tr_{\mathcal{R}}\Big[\frac{1}{6}\mathcal{D}^2E+\frac{1}{2}E^2+\frac{1}{6}RE+\frac{1}{12}\Omega_{\mu\nu}\Omega^{\mu\nu}\\
&+\frac{1}{30}\mathcal{D}^2R+\frac{1}{72}R^2-\frac{1}{180}R_{\mu\nu}R^{\mu\nu}+\frac{1}{180}R_{\mu\nu\rho\sigma}R^{\mu\nu\rho\sigma}\Big].
\end{split}
\ee 
where $\Omega_{\mu\nu}$ is the curvature of the appropriate covariant derivative operator.

\subsection{One-loop effective action} 
In the background field method, the effective action is defined by considering the quantum fluctuations around a fixed background field. The generating functional is given by 
\be 
\label{pathint}
e^{iW(\xi_0, J)}=\int \mathcal{D}\xi e^{i(S(\xi+\xi_0)+J\xi)}.
\ee 
So, the background field effective action is 
\be 
\Gamma[\xi_0, \xi]=W(\xi_0, J)-J\tilde{\xi}, 
\ee 
where 
\be 
\tilde{\xi}=\frac{\delta W(\xi_0,J)}{\delta J}.
\ee 
Under the Wick rotation, we go to the Euclidean signature and then define the Euclidean path integral analogous to (\ref{pathint}). In the absence of sources, we put $J=0$ and obtain 
\be
\begin{split} 
e^{-\Gamma(\xi_0)}&=\int \mathcal{D}\xi e^{-S(\xi+\xi_0)}\\
&=e^{-S(\xi_0)}\int\mathcal{D}\xi e^{-\int\xi\Delta\xi}.
\end{split} 
\ee 
Using the Gaussian integration formula 
\be 
\int\Pi_j dx_je^{-\frac{1}{2}x_a M_{ab}x_b}=[\textrm{det}(M)]^{-1/2}
\ee 
we find 
\be
\begin{split} 
e^{-\Gamma(\xi_0)}&=e^{-S(\xi_0)}[\textrm{det}(\Delta)]^{-1/2}.
\end{split} 
\ee 
where $\Delta$ is a generalized Laplace operator obtained after linearizing the action around the background field. We then have the one-loop effective action 
\be 
\Gamma_{1-loop}=\frac{1}{2}\textrm{log det}(\Delta).
\ee 
The relevant expression for the one-loop effective action is given using (\ref{logdet}) and (\ref{trace})
\be 
\Gamma_{1-loop}=-\frac{1}{2(4\pi)^2}\int_0^{\infty}dt~ t^{n-3}\sum_{n=0}^{\infty}\int d^4x \sqrt{g}~a^{\mathcal{R}}_n(E).
\ee 
In a renormalization scale, which is free of any mass scale, the running of the coupling constants is obtained from the logarithmically divergent part. This part is extracted from the third term in the above expansion, particularly 
\be 
\gamma=\frac{1}{(4\pi)^2}\int d^4x \sqrt{g}~a_2.
\ee 
which allows to write 
\be 
\Gamma^{\textrm{log}}_{1-loop}=-\frac{1}{2}\gamma\int_0^{\infty}\frac{dt}{t}=\gamma~\textrm{log}\frac{\delta}{\delta_0}.
\ee 
where we have regularized the integral of $t$ by introducing some cut-offs $t_{max}\approx 1/\delta^2$, $t_{min}\approx 1/\delta_0^2$. Therefore, we can write 
\be 
\frac{\partial \Gamma_{1-loop}^{\textrm{log}}}{\partial log \delta}=\gamma.
\ee 
\subsection{Gravitational sector}
The free part of the gauge fixed Lagrangian reads
\be  
\label{full1}
\begin{split}
\mathcal{L}_2&=-2H^{JkJ'K'}\tilde{\nabla}_{MJ'}\Omega^{~M}_{J~KK'}+4h^{JK}\tilde{\nabla}_{KJ'}\omega^{~J'}_{J}-\Omega^{JKMJ'}\Omega_{JKMJ'}+2\omega^{JJ'}\omega_{JJ'}
\end{split}
\ee 
As can be seen, it decouples into two sectors $(H,\Omega)$ and $(h,\omega)$. 
We will now proceed with the heat kernel computation in each of these sectors. To this end, we first rewrite the Lagrangian of the sector $(H,\Omega)$
\subsubsection{$(H,~\Omega)$ sector}
We write the Lagrangian in this sector as
\be 
\label{full2}
\mathcal{L}_{H\Omega}=\frac{1}{2}\begin{pmatrix}
H & \Omega 
\end{pmatrix}\begin{pmatrix}
0 & -2\tilde{\nabla} \\
-2\tilde{\nabla}^* & 2B
\end{pmatrix}\begin{pmatrix}
H \\
\Omega 
\end{pmatrix},
\ee
where $B$ is some endomorphism and is given by $B=-\epsilon^P_{~J}\epsilon^Q_{~K}\epsilon^R_{~M}\epsilon^{P'}_{~M'}$.\\~\\
Here 
\be \tilde{\nabla}:S^2_+\times S_+\times S_-\rightarrow S^2_+\times S_-\times S_-, ~~~~(\tilde{\nabla}\Omega)_{JKJ'K'}:=\tilde{\nabla}_{MJ'}\Omega^{~~M}_{JK~K'}, 
\ee 
is the chiral part of a Dirac type operator and 
\be 
\tilde{\nabla}^*:S^2_+\times S_-\times S_-\rightarrow S^2_+\times S_+\times S_-,~~~~
(\tilde{\nabla}^*H)_{JKMK'}:=\tilde{\nabla}^{J'}_M H_{JKJ'K'}
\ee 
is the adjoint operator. 
It is clear that the arising operator $D$ in (\ref{full2}) is of Dirac type and the standard method is to square this operator so that it becomes an operator of Laplace type. One can then use the heat kernel technology to compute the effective action. Squaring $D$ we have 
\be 
\label{full3}
D^2= \begin{pmatrix}
\tilde{\nabla}\tilde{\nabla}^* & -\tilde{\nabla}B \\
-B\tilde{\nabla}^* & \tilde{\nabla}^*\tilde{\nabla}+B^2\mathbb{1}
\end{pmatrix}.
\ee 
Let us compute the entries of the above matrix. We have 
\be
\begin{split}
(\tilde{\nabla}\tilde{\nabla}^{*}H)_{JKJ'M'}&=\tilde{\nabla}_{MJ'}\tilde{\nabla}^{MK'}H_{JKK'M'}\nonumber\\
&=-\frac{1}{2}\tilde{\nabla}^{NN'}\tilde{\nabla}_{NN'}H_{JKJ'M'}+2\tilde{\nabla}_{P'(J}\tilde{\nabla}_{P)}^{~P'}H^P_{~KJ'M'}+2\tilde{\nabla}_{P(J'}\tilde{\nabla}_{P')}^{~P}H^{~~~P'}_{JK~M'}\\
&=-\frac{1}{2}\tilde{\nabla}^{NN'}\tilde{\nabla}_{NN'}H_{JKM'J'}+2\square_{JP}H^{P}_{~KJ'M'}+2\square_{J'P'}H^{~~P'}_{JK~M'}\\
&=-\frac{1}{2}\tilde{\nabla}^{NN'}\tilde{\nabla}_{NN'}H_{JKM'J'}-12\Lambda H_{JKJ'M'}\\
&=-\frac{1}{2}\tilde{\nabla}^{NN'}\tilde{\nabla}_{NN'}H_{JKM'J'}
\end{split}
\ee 
where we have used the Lichnerowicz formula and used the fact that when one of the indices of the Weyl spinor is contracted, this gives zero. Thus the only non-zero contribution comes from the cosmological constant, but this vanishes in the setting where the cosmological constant is zero. Next, 
\be 
\begin{split}
(\tilde{\nabla}^*\tilde{\nabla}\Omega)_{JKNJ'}&=\tilde{\nabla}^{K'}_{~N}\tilde{\nabla}^{M}_{~K'}\Omega_{JKMJ'}\nonumber\\
&=-\frac{1}{2}\tilde{\nabla}^{NN'}\tilde{\nabla}_{NN'}\Omega_{JKNJ'}+2\tilde{\nabla}_{Q'(J}\tilde{\nabla}_{Q)}^{~Q'}\Omega^Q_{~KNJ'}+2\tilde{\nabla}_{Q(J'}\tilde{\nabla}_{Q')}^{~Q}\Omega^{~~~P'}_{JKN}\nonumber\\
&=-\frac{1}{2}\tilde{\nabla}^{NN'}\tilde{\nabla}_{NN'}\Omega_{JKNJ'}
\end{split}
\ee 
Thus the matrix we have is given by 
\be 
-2D^2=\begin{pmatrix}
\tilde{\nabla}^{\mu}\tilde{\nabla}_{\mu} & 2\tilde{\nabla}B \\
2B\tilde{\nabla}^* & \tilde{\nabla}^{\mu}\tilde{\nabla}_{\mu}-2B^2\mathbb{1}
\end{pmatrix}.
\ee 
We can rewrite $-2D^2$ as $D_{\mu}D^{\mu}-E$ where $D_{\mu}$ is a new matrix valued covariant derivative operator and $E$ is some endomorphism. The structure of $D_{\mu}$ that we recover is
\be 
D_{MM'}=\begin{pmatrix}
\tilde{\nabla}_{MM'} & B_{MM'} \\
B_{MM'} & \tilde{\nabla}_{MM'}
\end{pmatrix},
\ee
where we have suppressed the SL(2,C) indices in the spinorial object $B$, and 
\be 
E=\begin{pmatrix}
B^2\mathbb{1} & 0 \\
0 & 3B^2\mathbb{1}
\end{pmatrix}.
\ee
Multiplying $E$ with the Ricci scalar $R$, and then taking the trace with respect to the SL(2,C) indices, we get 
\be 
-\frac{1}{2.6}\textrm{tr}_{\mathcal{R}}(RE)=0.
\ee 
where we have used the fact that we are working for Ricci-flat background.
The curvature of the new connection $D_{MM'}$ is given by 
\be 
\mathcal{F}_{MM'NN'}=
\frac{1}{2}\begin{pmatrix}
\tilde{\nabla}_{[MM'}\tilde{\nabla}_{NN']}+B_{[MM'}B_{NN']} & -\tilde{\nabla}_{[MM'}B_{NN']}-B_{[MM'}\tilde{\nabla}_{NN']} \\
-\tilde{\nabla}_{[MM'}B_{NN']}-B_{[MM'}\tilde{\nabla}_{NN']} & \tilde{\nabla}_{[MM'}\tilde{\nabla}_{NN']}+B_{[MM'}B_{NN']}
\end{pmatrix}.
\ee 
Let us compute the components of the above matrix. The diagonal parts contribute
\be
\begin{split}
\label{sol1}
\tilde{\nabla}_{[MM'}\tilde{\nabla}_{NN']}+B_{[MM'}B_{NN']}&=F^{AB}_{~~~MM'NN'}\Sigma_{AB}+B_{[MM'}B_{NN']}\\
&=\psi^{AB}_{~~MN}\epsilon_{M'N'}\Sigma_{AB}+\epsilon^R_{~M}\epsilon^{P'}_{~M'}\epsilon^S_{~N}\epsilon^{S'}_{~N'}-\epsilon^R_{~N}\epsilon^{P'}_{~N'}\epsilon^S_{~M}\epsilon^{S'}_{~M'}.
\end{split}
\ee 
We do not need to explicitly compute the off-diagonal elements as these vanish. This is because these elements are both symmetric and antisymmetric with respect to $MM'$ and $NN'$ and this is why there are vanishing. 
We need to take the trace of the square of the curvature because this is what is there in the heat kernel coefficient. It is straightforward to work this out. To compute explicitly, we have 
\be 
\begin{split}
&2\textrm{tr}_{\mathcal{R}}\Big[\psi^{AB}_{~~MN}\epsilon_{M'N'}\Sigma_{AB}\psi^{CDMN}\epsilon^{M'N'}\Sigma_{CD}\\&+2\psi^{AB}_{~~MN}\epsilon_{M'N'}\Sigma_{AB}(\epsilon^{RM}\epsilon^{P'M'}\epsilon^{SN}\epsilon^{S'N'}-\epsilon^{RN}\epsilon^{P'N'}\epsilon^{SM}\epsilon^{S'M'})\Big],
\end{split}
\ee
where we have neglected the square terms coming from the epsilon metrics because these vanish. The second term on the above line vanishes because of the symmetry property of the Weyl spinor. Thus we are left with the first term. Evaluating the trace, we get
\be 
\begin{split}
\textrm{tr}_{\mathcal{R}}\Big[\frac{1}{12}\mathcal{F}_{\mu\nu}\mathcal{F}^{\mu\nu}\Big]&=\frac{1}{12.4}\textrm{tr}_{\mathcal{R}}[4\psi^{ABMN}\psi_{CDMN}]
\nonumber\\
&=\frac{1}{12}\psi^{ABMN}\psi_{ABMN}.
\end{split}
\ee 
The next object to compute is $E^2$. We have 
\be 
E^{'2}=\Big(-\frac{1}{2}E\Big)^2=\begin{pmatrix}
(\frac{1}{2}B^2)^2 & 0 \\
0 & (\frac{3}{2}B^2)^2
\end{pmatrix}
\ee 
Evaluating the trace of this operator, we get 
\be 
\frac{1}{2}\textrm{tr}_{\mathcal{R}}(E^2)=0
\ee 
\subsubsection{$(h,\omega)$ sector}
The kinetic part of the Lagrangian in this sector is analogous to what we found in the $(H,\Omega)$ sector. The only difference that arises is because of factors of 2 which we absorb in the definition of the fields $(h,\omega)$. Let us rewrite the free part of the Lagrangian as follows 
\be 
\label{full6}
\mathcal{L}_{h\omega}=\frac{1}{2}\begin{pmatrix}
h & \omega 
\end{pmatrix}\begin{pmatrix}
0 & 4\tilde{\nabla}^* \\
4\tilde{\nabla} & 4\tilde{B} 
\end{pmatrix}\begin{pmatrix}
h \\
\omega 
\end{pmatrix}.
\ee 
Here 
\be 
\tilde{\nabla}^*: S_-\rightarrow S_+, ~~~~(\tilde{\nabla}\omega)_{MJ}:=\tilde{\nabla}_{MJ'}\omega^{J'}_{~J}, 
\ee 
is the chiral part of a Dirac type operator and 
\be 
\tilde{\nabla}:S_+\rightarrow S_-,~~~~
(\tilde{\nabla}^*h)_{KJ'}:=\tilde{\nabla}^{J}_{~J'} h_{JK}
\ee 
is the adjoint operator. 
It is transparent that the arising operator $\tilde{D}$ in (\ref{full6}) is of Dirac type. The computation in this sector exactly mimicks the one in the previous. We do not repeat this computation here but outline the main results of interest. 
\be 
\label{full7}
\tilde{D}^2=\begin{pmatrix}
-2\tilde{\nabla}^{\mu}\tilde{\nabla}_{\mu}& 4\tilde{\nabla}^*\tilde{B} \\
4\tilde{B}\tilde{\nabla} & -2\tilde{\nabla}^{\mu}\tilde{\nabla}_{\mu}+4\tilde{B}^2\mathbb{1}
\end{pmatrix}.
\ee 
We can rewrite $-\frac{1}{2}\tilde{D}^2$ as $\tilde{D}_{\mu}\tilde{D}^{\mu}-\tilde{E}$ where $\tilde{D}_{\mu}$ is a new matrix valued covariant derivative operator and $\tilde{E}$ is some endomorphism. 
\be 
\tilde{D}_{MM'}=\begin{pmatrix}
\tilde{\nabla}_{MM'} & -\tilde{B}_{MM'} \\
-\tilde{B}_{MM'} & \tilde{\nabla}_{MM'}
\end{pmatrix},
\ee
and 
\be 
\tilde{E}=\begin{pmatrix}
-B^2\mathbb{1} & 0 \\
0 &-3B^2\mathbb{1}
\end{pmatrix}.
\ee 
We can directly write the curvature of the new connection which is given by
\be 
\mathcal{\tilde{F}}_{MM'NN'}=-2\begin{pmatrix}
\tilde{\nabla}_{[MM'}\tilde{\nabla}_{NN']}+\tilde{B}_{[MM'}\tilde{B}_{NN']} & -\tilde{\nabla}_{[MM'}\tilde{B}_{NN']}-\tilde{B}_{[MM'}\tilde{\nabla}_{NN']} \\
-\tilde{\nabla}_{[MM'}\tilde{B}_{NN']}-\tilde{B}_{[MM'}\tilde{\nabla}_{NN']} & \tilde{\nabla}_{[MM'}\tilde{\nabla}_{NN']}+\tilde{B}_{[MM'}\tilde{B}_{NN']}
\end{pmatrix}.
\ee 
The trace of the square of above object is already computed explicitly in the previous sector, upto some numerical factors. Restoring the correct factors, we quote the result here
\be 
\begin{split}
\textrm{tr}_{\mathcal{R}}\Big[\frac{1}{12}\mathcal{\tilde{F}}_{\mu\nu}\mathcal{\tilde{F}}^{\mu\nu}\Big]&=\frac{4}{12}\textrm{tr}_R[4\psi^{ABMN}\psi_{CDMN}]
\nonumber\\
&=\frac{4}{3}\psi^{ABMN}\psi_{ABMN}.
\end{split}
\ee 
The trace of the endomorphism $E$ is given by 
\be
\frac{2}{6}\textrm{tr}_{\mathcal{R}}(RE)=0.
\ee 
The next object to compute is $\tilde{E}^2$. We have 
\be 
\tilde{E}^2=\begin{pmatrix}
(\tilde{B}^2)^2 & 0 \\
0 & (3\tilde{B}^2)^2
\end{pmatrix}.
\ee 
However, the trace of $\tilde{E}^2$ is zero.
We convert the result of the square of Weyl spinors in terms of $R^2$, $R_{\mu\nu}^2$ and $R_{\mu\nu\rho\sigma}^2$. However, both $R^2$ and $R_{\mu\nu}$ vanish and the heat kernel coefficient in the gravitational sector is given by 
\be 
a_{2bosonic}^{\mathcal{R}}(E)=\frac{64}{45}R_{\mu\nu\rho\sigma}R^{\mu\nu\rho\sigma}.
\ee 

\subsection{Ghost sector}
Let us now return to the ghost sector.
The Lagrangian in (\ref{ghost3!}) is not diagonal in the  fields. It has mixed terms which arise from the coupling of diffeomorphism and Lorentz ghosts. To do a heat kernel computation, it is necessary to obtain a diagonalised Lagrangian such that the differential operator which arises is of Laplace type. Thus, we now proceed to diagonalize it. Let us first give a outline of the diagonalization procedure and then we will implement it in our problem. Consider a differential operator of the form 
\be 
\partial^\mu \partial_\mu \mathds{1} + \partial_\mu A^\mu + B,
\ee 
where $A^{\mu}$ and $B$ are matrices in some matrix space. This operator can in turn be written as a matrix which then acts on the column vector comprising of fields $b,c$. The conjugate fields $\bar{b},\bar{c}$ are contracted from the left and this produces the Lagrangian. It is then possible to absorb the linear parts of the differential operator by a suitable redefinition such that it results in an operator of the form 
\be 
\mathcal{D}^{\mu} \mathcal{D}_{\mu} + E,
\ee 
where $E$ is some endomorphism on the field space and $\mathcal{D}_{\mu}$ is an appropriate covariant derivative operator. It is customary to first write the Lagrangian in (\ref{ghost3!}) in a matrix form, so that one can diagonalize the matrix by suitable field re-definitions. We have
\be 
\label{ghost3}
\mathcal{L}_{ghost}&=-\begin{pmatrix}
    \bar{b}^{JK} & \bar{c}^{BB'}  
  \end{pmatrix}
  \begin{pmatrix}
   X &  Y\\
   Z & X
  \end{pmatrix}
  \begin{pmatrix}
    b_{JK} \\
    c_{BB'}\\
   \end{pmatrix},
\ee 
where 
\be 
\begin{split} 
X&=\tilde{\nabla}_{MM'}\tilde{\nabla}^{MM'},\\
Y&=\frac{1}{2}\psi^{MB}_{~~~JK}\tilde{\nabla}^{~B'}_{M},\\
Z&=-\tilde{\nabla}^{K}_{~B'}\epsilon^{J}_{~B}.\\
\end{split} 
\ee 
where in writing $Y$, we used the Bianchi identity $\nabla_{MB'}\psi^{M}_{~~BJK}=0$ and got rid of the other term.
The matrix in (\ref{ghost3}) can be split in the following way \be
\begin{split}
\label{matrix}
 \begin{pmatrix}
   X &  Y\\
   Z & X
  \end{pmatrix}&=
\tilde{\nabla}_{MM'}\tilde{\nabla}^{MM'}\begin{pmatrix}
   1 &  0\\
   0 & 1
  \end{pmatrix}+
 \begin{pmatrix}
   0 & \frac{1}{2}\epsilon^{M'B'} \psi^{MB}_{~~~JK}\\
  -\epsilon^{J}_{~B}\epsilon^{KM}\epsilon^{M'}_{~B'} & 0
  \end{pmatrix}\tilde{\nabla}_{MM'} \\~\\&+
  \tilde{\nabla}^{MM'}\begin{pmatrix}
    0 & \frac{1}{2}\epsilon_{M'B'} \psi^{B}_{~~MJK} \\
  -\epsilon^{J}_{~B}\epsilon^{K}_{~M}\epsilon_{~M'B'} & 0
  \end{pmatrix}.
  \end{split}
  \ee 
This can now be re-written by absorbing the first order derivative parts and expressing it in the form of a Laplace type operator 
\be 
\begin{pmatrix}
   X &  Y\\
   Z & X
  \end{pmatrix}&=\mathcal{D}^{MM'}\mathcal{D}_{MM'}+E,
\ee 
where we define the new connection
\be 
\label{connnew}
\mathcal{D}^{MM'}=\begin{pmatrix}
   \tilde{\nabla}^{MM'} &  \frac{1}{2}\epsilon^{M'B'} \psi^{MB}_{~~~JK} \\~\\
  \epsilon^{J}_{~B}\epsilon^{KM}\epsilon^{M'}_{~B'}  & \tilde{\nabla}^{MM'}
  \end{pmatrix}
 \ee 
and 
\be
E=-\begin{pmatrix}
   \psi^{KJ}_{~~~JK} &  0 \\~\\
  0 & \psi^{KJ}_{~~~JK} 
  \end{pmatrix}.
 \ee 
Upon index contractions, the self dual part of the Weyl tensor
vanishes and therefore we can set $E=0$ henceforth. 
We thus have the relevant Laplace type operator, which in this case with $E=0$ is 
\be 
\Delta :=\mathcal{D}_{MM'}\mathcal{D}^{MM'}.
\ee
with the new connection defined in (\ref{connnew}).

We define $\Omega_{\mu\nu}$ as
\be 
\begin{split}
\Omega_{\mu\nu}&=[\mathcal{D}_{\mu},\mathcal{D}_{\nu}]\\
&=\begin{pmatrix}
   \tilde{\nabla}_{\mu} &   Y_{1\mu}\\
   Y_{2\mu} & \tilde{\nabla}_{\mu}
  \end{pmatrix}
  \begin{pmatrix}
   \tilde{\nabla}_{\nu} &   Y_{1\nu}\\
   Y_{2\nu} & \tilde{\nabla}_{\nu}
  \end{pmatrix}
  -\begin{pmatrix}
   \tilde{\nabla}_{\nu} &   Y_{1\nu}\\
   Y_{2\nu} & \tilde{\nabla}_{\nu}
  \end{pmatrix}
  \begin{pmatrix}
   \tilde{\nabla}_{\mu} &   Y_{1\mu}\\
   Y_{2\mu} & \tilde{\nabla}_{\mu}
  \end{pmatrix}\\
&=\begin{pmatrix}
   \tilde{\nabla}_{[\mu}\tilde{\nabla}_{\nu]}+Y_{1[\mu}Y_{2\nu]} &  \tilde{\nabla}_{[\mu}Y_{1\nu]} +Y_{1[\mu}\tilde{\nabla}_{\nu]}\\~\\
   Y_{2[\mu}\tilde{\nabla}_{\nu]}+\tilde{\nabla}_{[\mu}Y_{2\nu]} & Y_{2[\mu}Y_{1\nu]}+\tilde{\nabla}_{[\mu}\tilde{\nabla}_{\nu]}
  \end{pmatrix}.
 \end{split}
\ee
where $Y^{MM'}_{1}=\frac{1}{2}\epsilon^{M'B'}\psi^{MB}_{~~~JK}$ and $Y^{MM'}_{2}=\epsilon^{J}_{~B}\epsilon^{KM}\epsilon^{M'}_{~B'}$.
Let us compute the matrix elements of $\Omega_{\mu\nu}$ using spinor notations, which will yield simpler expressions. The diagonal elements are equal to the commutator of the total covariant derivative. This commutator can act on the Lorentz ghost which has just the internal indices or the diffeomorphism ghost which has only the spacetime indices. Whatever the ghost field, the action of the commutator can only be proportional to the curvature of the spin connection. This is because the curvature of the spin connection is related to the curvature of the Levi-civita connection via the tetrad one-forms and thus they are equivalent descriptions of the same entity. Further, the background field equation expresses the curvature in terms of the Weyl spinor. We already arrived at a simple form of the curvature in spinor notations when expressed in terms of the Weyl spinor. Thus, overall the diagonal part of $\Omega_{\mu\nu}$ in spinor notations contributes  
\be 
\begin{split} 
\label{matrixelements}
\tilde{\nabla}_{[MM'}\tilde{\nabla}_{NN']}+Y_{1[MM'}Y_{2NN']}
&=\epsilon_{M'N'}\psi^{AB}_{~~MN}\Sigma_{AB},
\end{split} 
\ee 
where we used the fact that when two of the indices of the Weyl spinor are contracted, it vanishes, i.e $\psi^{J}_{~MJK}=0$. Thus the contribution from the second term of the left hand side above is zero. Let us now compute the off-diagonal parts. The lower off-diagonal element when evaluated in spinor notations give 
\be
\begin{split} 
\tilde{\nabla}_{[\mu}Y_{2\nu]} +Y_{2[\mu}\tilde{\nabla}_{\nu]}&=\tilde{\nabla}_{MM'}\epsilon^{J}_{~B}\epsilon^{K}_{~N}\epsilon_{N'B'}-\tilde{\nabla}_{NN'}\epsilon^{J}_{~B}\epsilon^{K}_{~M}\epsilon_{M'B'}\\&+\epsilon^{J}_{~B}\epsilon^{K}_{~M}\epsilon_{M'B'}\tilde{\nabla}_{NN'}-\epsilon^{J}_{~B}\epsilon^{K}_{~N}\epsilon_{N'B'}\tilde{\nabla}_{MM'}\\
&=0.
\end{split} 
\ee 
Let us also compute the upper right off-diagonal part. This is given by 
\be
\begin{split} 
\tilde{\nabla}_{[\mu}Y_{1\nu]} +Y_{1[\mu}\tilde{\nabla}_{\nu]}&=\epsilon_{N'C'}\nabla_{MM'}\psi^{JK}_{~~NC}-\epsilon_{M'C'}\nabla_{NN'}\psi^{JK}_{~~MC}\\&+\psi^{JK}_{~~MC}\epsilon_{M'C'}\tilde{\nabla}_{NN'}-\psi^{JK}_{~~NC}\epsilon_{N'C'}\tilde{\nabla}_{MM'}.
\end{split} 
\ee 
We now need to compute $\Omega^2$. Let us first see the structure of the off-diagonal elements which arise in it. Indeed, it is easy to check that the lower left off-diagonal element must vanish and the upper left will be given by
\be 
\begin{split} 
&\psi^{MN}_{~~~AB}\epsilon^{M'N'}\Big(\epsilon_{N'C'}\nabla_{MM'}\psi^{JK}_{~~NC}-\epsilon_{M'C'}\nabla_{NN'}\psi^{JK}_{~~MC}\\&+\psi^{JK}_{~~MC}\epsilon_{M'C'}\tilde{\nabla}_{NN'}-\psi^{JK}_{~~NC}\epsilon_{N'C'}\tilde{\nabla}_{MM'}\Big).
\end{split}
\ee 
Due to index contractions of the Weyl spinor, the above term vanishes. Then, the square of the curvature is given entirely by diagonal terms, which are however equal and is given by 
\be 
\begin{split} 
\Omega_{MM'NN'}\Omega^{MM'NN'}&=\begin{pmatrix}
  2\psi_{ABMN}\psi^{CDMN}  &  0 \\~\\
 0 & 2\psi_{ABMN}\psi^{CDMN}
  \end{pmatrix}.
\end{split}
\ee
Substituting this in the heat kernel formula (\ref{coeff}), we have the relevant for the heat kernel coefficient as
\be
\label{coeff2}
\begin{split} 
\frac{1}{12}\textrm{tr}_{\mathcal{R}}[\Omega_{\mu\nu}\Omega^{\mu\nu}]&=\frac{1}{3}\psi_{MNPS}\psi^{MNPS}
\end{split}
\ee 
Let us now rewrite the above as a linear combination of the curvature squares. Doing so, we have for the heat kernel coefficient
\be 
a_{2ghost}^{\mathcal{R}}(E)=\frac{61}{180}R_{\mu\nu\rho\sigma}R^{\mu\nu\rho\sigma}
\ee 
\subsection{Final result and comparison with metric gravity} 
The final result for the heat kernel coefficient is given by 
\be 
a_{2bosonic}^{\mathcal{R}}-2a_{2ghost}^{\mathcal{R}}=\frac{67}{90}R_{\mu\nu\rho\sigma}R^{\mu\nu\rho\sigma}
\ee 
Thus, 
\be 
\gamma_{chiral-EC}=\frac{1}{(4\pi)^2}\int d^4x\sqrt{g}\Bigg(\frac{67}{90}R_{\mu\nu\rho\sigma}R^{\mu\nu\rho\sigma}\Bigg)
\ee 
The analogous result in metric gravity is 
\be 
\gamma_{metric}=\frac{1}{(4\pi)^2}\int d^4x\sqrt{g}\Bigg(\frac{53}{45}R_{\mu\nu\rho\sigma}R^{\mu\nu\rho\sigma}\Bigg)
\ee 
Although the above results are total derivatives, we notice that the coefficient of $R_{\mu\nu\rho\sigma}^2$ divergences changes. This alreads tells us that the quantum theory for the chiral Einstein-Cartan gravity is different than metric GR. It would be interesting to see what happens when we couple chiral Einstein-Cartan gravity to matter or deal with a non-zero cosmological constant. 
\section{Discussion}
In this paper, we studied the chiral Einstein-Cartan action on an arbitrary Einstein background for zero cosmological constant. We developed the gauge-fixing procedure using the BRST formalism. The action and the fields appearing in the first order formalism are all expressed in spinor notations to conveniently perform the gauge fixing. The diffeomorphism and Lorentz gauge fixing, when done on an Einstein background, remarkably lead to decoupling of the free part of the Lagrangian into two sectors, mimicking the result of its flat version \cite{Krasnov:2020bqr}. The gauge fixing fermion that we chose to work with is similar to the one in its flat space version, except that now it is non-linear in the fields. Nevertheless, this non-linearity does not cause us problems and the bosonic part decouples nicely into two sectors, namely $(H,\Omega)$ and $(h,\omega)$.\\However, the novelty of this result is the appearance of a new quadratic term for the tetrad/metric field. This new term does not affect the gauge fixing procedure and does not complicate calculations.
Let us now comment on the BRST complex in this first-order formalism. The complex is so designed that the diffeomorphism ghosts appear in the transformation of the Lorentz ghost. As a result, it leads to the coupling of diffeomorphism and Lorentz ghosts in the ghost Lagrangian. Also, we massaged the transformation for the tetrad and the spin-connection in such a way that the diffeomorphism transform does not contain the derivatives of the vector field which generates it. For the spin-connection, the vector field is inserted into the curvature of the spin-connection and this is corrected by a gauge transformation.
Using the gauge fixed Lagrangian, we computed the heat kernel coefficients appearing in the one-loop effective action by computing the determinant of the operators arising in the bosonic and the ghost sectors. In the bosonic sector, this operator is a version of the Dirac operator, but with background covariant derivatives everywhere. We squared it to get the relevant Laplace type operator, from which we extracted the heat kernel coefficient following the standard technology. In the ghost sector, the operator arising in the Lagrangian is not of Laplace type. By suitable field redefinitions, we converted it into a Laplace-type operator. We found that the endomorphism part vanishes because what appears here is the Weyl spinor with its indices contracted. We found that the arising one-loop divergences are proportional to the $R_{\mu\nu\rho\lambda}^2$ term. However, the coefficient of this term is different from the corresponding result in metric GR. This is not unexpected because, although on-shell, the chiral formulation is equivalent to metric GR, in an off-shell calculation there is no guarantee for such a similarity. Indeed, this is what we found by doing a one-loop computation. This is to be distinguished from the result of the pure connection formalism, achieved in \cite{Groh:2013oaa}, where the sign in front of the $R_{\mu\nu\rho\lambda}^2$ term was different and there was a close relation with the metric GR. Our result is important in the context of quantization of different formulations of gravity and in understanding whether they lead to different quantum theories. 
\\~\\
It is possible to understand the renormalization flow of any such equivalent formulation and search for a UV-fixed point in the ultraviolet. Indeed, this has been done in the pure-connection formulation. In our present work, we do not analyze the flow but instead only compute the one-loop effective action. A more general study of RG flow can be done in a setting where the cosmological constant is non-zero and the background is completely arbitrary. Also, a two-loop computation is possible in such a context. We leave this for future research.  
\\~\\
There are other interesting calculations that can be done in the present context. It is possible to couple fermions to the chiral Einstein-Cartan action, which would lead to a nonzero torsion. Then one can study the one-loop divergences that arise in such a setting.
In this paper, we developed the ghost Lagrangian for an arbitrary Einstein background. It is possible to take its flat space limit by replacing covariant derivatives with ordinary partial derivatives everywhere. This can be used to compute ghost diagrams for loop level scattering amplitudes. We hope to return to such questions in a future publication.
\section{Acknowledgements}
I am grateful to Kirill Krasnov for suggesting this interesting work and also for various discussions. I am also grateful to K P Yogendran for carefully reading the manuscript and for his insightful comments on the work. 
\section{Appendix: BRST closure}
\subsection{Diffeomorphism ghost} 
Let us start with the second transformation in (\ref{brstn}), which is that of the diffeomorphism ghost. We have 
\be 
\begin{split} 
s^2c_{\mu}&=sc^{\lambda}\nabla^T_{\lambda}c_{\mu}-c^{\lambda}\nabla^T_{\lambda}sc^{\mu}\\
&=c^{\gamma}\nabla_{\gamma}c^{\lambda}\nabla_{\lambda}c^{\mu}-c^{\lambda}\nabla_{\lambda}c^{\nu}\nabla_{\nu}c^{\mu}-c^{\lambda}c^{\nu}\nabla_{\lambda}\nabla_{\nu}c^{\mu}\\
&=-c^{\lambda}c^{\nu}\nabla_{\lambda}\nabla_{\nu}c^{\mu}\\
&=-\frac{1}{2}(c^{\lambda}c^{\nu}-c^{\nu}c^{\lambda})\nabla_{\lambda}\nabla_{\nu}c^{\mu}\\
&=-\frac{1}{2}c^{\lambda}c^{\nu}[\nabla_{\lambda},\nabla_{\nu}]c^{\mu}\\&=-\frac{1}{2}c^{\lambda}c^{\nu}R^{\mu}_{~\alpha\nu\lambda}c^{\alpha}\\
&=-\frac{1}{2}c^{\alpha}c^{\lambda}c^{\nu}R^{\mu}_{~\alpha\nu\lambda}\\
&=-\frac{1}{6}c^{\alpha}c^{\lambda}c^{\nu}(R^{\mu}_{~\alpha\nu\lambda}+R^{\mu}_{~\nu\lambda\alpha}+R^{\mu}_{~\lambda\alpha\nu})=0.
\end{split}
\ee 
where we have used the sign carefully to take care of the anti-commutativity of the ghosts. In the third line of the above, we used this property to write the single term as a combination of two terms, which then becomes the commutator of the covariant derivatives. Finally, we express this commutator in terms of the Riemann tensor and used the symmetry properties of it to deduce that the expression vanishes due to the first Bianchi identity.
\subsection{Tetrad}
For the transformation of the tetrad, we have 
\be 
\label{tetrad}
\begin{split}
s^2h^i_{\mu}&=sc^{\nu}\nabla_{\nu}h^{i}_{\mu}(x)-c^{\nu}\nabla_{\nu}sh^{i}_{\mu}(x)+sh^{i}_{\nu}(x)\nabla_{\mu}c^{\nu}+h^{i}_{\nu}(x)\nabla_{\mu}sc^{\nu}\\
&=c_{\gamma}\nabla^{\gamma}c^{\delta}\nabla_{\delta} h^i_{\mu}-c^{\nu}\nabla_{\nu}c^{\delta}\nabla_{\delta}h^i_{\mu}-c^{\nu}c^{\delta}\nabla_{\delta}\nabla_{\nu}h^i_{\mu}\\&-c^{\gamma}\nabla_{\gamma}h^i_{\nu}\nabla_{\mu}c^{\nu}-c^{\gamma}h^i_{\nu}\nabla_{\gamma}\nabla_{\mu}c^{\nu}+c^{\gamma}\nabla_{\gamma}h^i_{\nu}\nabla_{\mu}c^{\nu}+h^i_{\gamma}\nabla_{\nu}c^{\gamma}\nabla_{\mu}c^{\nu}\\&+h^i_{\nu}\nabla_{\mu}c^{\gamma}\nabla_{\gamma}c^{\nu}+h^i_{\nu}c^{\gamma}\nabla_{\mu}\nabla_{\gamma}c^{\nu}\\
&=h^i_{\nu}c^{\gamma}\nabla_{\mu}\nabla_{\gamma}c^{\nu}-c^{\gamma}h^i_{\nu}\nabla_{\gamma}\nabla_{\mu}c^{\nu}-c^{\gamma}c^{\delta}\nabla_{\delta}\nabla_{\gamma}h^i_{\mu}\\
&=h^i_{\nu}c^{\gamma}[\nabla_{\mu},\nabla_{\gamma}]c^{\nu}+\frac{1}{2}c^{\delta}c^{\gamma}[\nabla_{\delta},\nabla_{\gamma}]h^i_{\mu}\\
&=h^i_{\nu}c^{\gamma} R^{\nu}_{~\alpha\gamma\mu}c^{\alpha}+\frac{1}{2}c^{\delta}c^{\gamma}R^{\beta}_{~\mu\gamma\delta}h^i_{\beta}
=h^i_{\nu}c^{\gamma}c^{\alpha} R^{\nu}_{~\alpha\gamma\mu}+\frac{1}{2}c^{\delta}c^{\gamma}R^{\beta}_{~\mu\gamma\delta}h^i_{\beta}\\
&=\frac{1}{2}h^i_{\nu}(c^{\gamma}c^{\alpha}-c^{\alpha}c^{\gamma})R^{\nu}_{~\alpha\gamma\mu}+\frac{1}{2}c^{\delta}c^{\gamma}R^{\beta}_{~\mu\gamma\delta}h^i_{\beta}\\
&=\frac{1}{2}h^i_{\nu}c^{\gamma}c^{\alpha} R^{\nu}_{~\alpha\gamma\mu}+\frac{1}{2}h^i_{\nu}c^{\gamma}c^{\alpha} R^{\nu}_{~\gamma\mu\alpha}+\frac{1}{2}c^{\delta}c^{\gamma}R^{\beta}_{~\mu\gamma\delta}h^i_{\beta}\\
&=\frac{1}{2}h^i_{\nu}c^{\gamma}c^{\alpha} (R^{\nu}_{~\alpha\gamma\mu}+R^{\nu}_{~\gamma\mu\alpha}+R^{\nu}_{~\mu\alpha\gamma})=0.
\end{split}
\ee 
In the second line of the above, many terms get cancelled and we are left with only three terms. These terms are rearranged in the form of the commutator of covariant derivatives which are further expressed again in terms of the Riemann tensor.
\subsection{Lorentz ghost} 
For the transformation of the Lorentz ghost field, we find 
\be 
\label{lg}
\begin{split}
s^2b^{ij}&=-\frac{1}{2}s[b,b]^{ij}+\frac{1}{2}s(c^{\mu}c^{\nu}F^{ij}_{~~\mu\nu})\\
&=-sb^{i}_{~k}b^{kj}+b^{i}_{~k}sb^{kj}+\frac{1}{2}sc^{\mu}c^{\nu}F^{ij}_{~~\mu\nu}\\
&-\frac{1}{2}c^{\mu}sc^{\nu}F^{ij}_{~~\mu\nu}+\frac{1}{2}c^{\mu}c^{\nu}sF^{ij}_{~~\mu\nu}\\
&=b^{i}_{~m}b^m_{~k}b^{kj}-\frac{1}{2}c^{\mu}c^{\nu}F^{i}_{~k\mu\nu}b^{kj}-b^i_{~k}b^{k}_{~m}b^{mj}\\
&+\frac{1}{2}b^i_{~k}c^{\mu}c^{\nu}F^{kj}_{~~\mu\nu}+\frac{1}{2}c^{\lambda}\nabla^T_{\lambda}c^{\mu}c^{\nu}F^{ij}_{~~\mu\nu}
-\frac{1}{2}c^{\mu}c^{\lambda}\nabla^T_{\lambda}c^{\nu}F^{ij}_{~~\mu\nu}\\
&+\frac{1}{2}c^{\mu}c^{\nu}c^{\lambda}\nabla^T_{\lambda}F^{ij}_{~~\mu\nu}+\frac{1}{2}c^{\mu}c^{\nu}F^{ij}_{~~\mu\lambda}\nabla^T_{\nu}c^{\lambda}+\frac{1}{2}c^{\mu}c^{\nu}F^{ij}_{~~\lambda\nu}\nabla^T_{\mu}c^{\lambda}+\frac{1}{2}c^{\mu}c^{\nu}[F_{\mu\nu},b]^{ij}\\
&=\frac{1}{2}c^{\mu}c^{\nu}c^{\lambda}\nabla^T_{\lambda}F^{ij}_{~~\mu\nu}
=\frac{1}{6}c^{\mu}c^{\nu}c^{\lambda}(\nabla^T_{\lambda}F^{ij}_{~~\mu\nu}+\nabla^T_{\mu}F^{ij}_{~~\nu\lambda}+\nabla^T_{\nu}F^{ij}_{~~\lambda\mu})\\
&=0.
\end{split}
\ee
where in the last line, we used the Bianchi identity and the fact that the Lorentz ghost fields anti commute and hence can be appropriately permuted to give rise to three such terms.

\subsection{Connection}
Let us now check the closure for the connection field. We have the following. \\
\be
\label{connection}
\begin{split}
s^2\omega^{ij}_{~\mu}&=\nabla^T_{\mu}sb^{ij}+sc^{\nu}F^{ij}_{~~\nu\mu}-c^{\nu}sF^{ij}_{~~\nu\mu}\\
&=-\nabla^T_{\mu}(b^i_{~k}b^{kj})+\frac{1}{2}\nabla^T_{\mu}(c^{\mu}c^{\nu}F^{ij}_{~~\mu\nu})+c^k\nabla^T_{k}c^{\nu}F^{ij}_{~~\nu\mu}-c^{\nu}[F_{\nu\mu},b]^{ij}\\
&-c^{\nu}c^{\lambda}\nabla^T_{\lambda}F^{ij}_{~~\nu\mu}-c^{\nu}F^{ij}_{~~\nu\lambda}\nabla^T_{\mu}c^{\lambda}-c^{\nu}F^{ij}_{~~\lambda\mu}\nabla^T_{\nu}c^{\lambda}\\
&=-\nabla^T_{\mu}(b^i_{~k}b^{kj})-\frac{1}{2}c^{\nu}\nabla^T_{\mu}c^{\mu}F^{ij}_{~~\mu\nu}+\frac{1}{2}c^{\mu}\nabla^T_{\mu}c^{\nu}F^{ij}_{~~\mu\nu}\\&+\frac{1}{2}c^{\mu}c^{\nu}\nabla^T_{\mu}F^{ij}_{~~\mu\nu}+c^k\nabla^T_{k}c^{\nu}F^{ij}_{~~\nu\mu}-c^{\nu}c^{\lambda}\nabla^T_{\lambda}F^{ij}_{~~\nu\mu}\\&-c^{\nu}F^{ij}_{~~\nu\lambda}\nabla^T_{\mu}c^{\lambda}-c^{\nu}F^{ij}_{~~\lambda\mu}\nabla^T_{\nu}c^{\lambda}-c^{\nu}[F_{\nu\mu},b]^{ij}\\
&=-\nabla^T_{\mu}(b^i_{~k}b^{kj})-c^{\nu}[F_{\nu\mu},b]^{ij}\\
&-\frac{1}{2}c^{\nu}c^{\lambda}(\nabla^T_{\lambda}F^{ij}_{~~\nu\mu}+\nabla^T_{\nu}F^{ij}_{~~\mu\lambda}+\nabla^T_{\mu}F^{ij}_{~~\lambda\nu})\\
&=0.
\end{split}
\ee 
where in the above, we used the Bianchi identity for the curvature, interchanged dummy indices and used the anti-commutative nature of the ghost fields.
This completes the check of the nilpotency of the BRST operator.


\begin{thebibliography}{99}

\bibitem{Palatini}
Deduzione invariantiva delle equazioni gravitazionali dal principio di Hamilton
A. Palatini
Rend.Circ.Mat.Palermo 43 (1919) 203-212


\bibitem{tHooft:1974toh}
G.~'t Hooft and M.~J.~G.~Veltman,
``One loop divergencies in the theory of gravitation,''
Ann. Inst. H. Poincare Phys. Theor. A \textbf{20} (1974), 69-94

\bibitem{Goroff:1985sz}
M.~H.~Goroff and A.~Sagnotti,
``QUANTUM GRAVITY AT TWO LOOPS,''
Phys. Lett. B \textbf{160} (1985), 81-86
doi:10.1016/0370-2693(85)91470-4

\bibitem{Goroff:1985th}
M.~H.~Goroff and A.~Sagnotti,
``The Ultraviolet Behavior of Einstein Gravity,''
Nucl. Phys. B \textbf{266} (1986), 709-736
doi:10.1016/0550-3213(86)90193-8

\bibitem{Gronwald:1995em}
F.~Gronwald and F.~W.~Hehl,
``On the gauge aspects of gravity,''
[arXiv:gr-qc/9602013 [gr-qc]].

\bibitem{Frolov:2009wu}
A.~M.~Frolov, N.~Kiriushcheva and S.~V.~Kuzmin,
``The Hamiltonian Formulation of Tetrad Gravity: Three-Dimensional Case,''
Grav. Cosmol. \textbf{16} (2010), 181-194
doi:10.1134/S0202289310030011
[arXiv:0902.0856 [gr-qc]].


\bibitem{Kiriushcheva:2009tg}
N.~Kiriushcheva and S.~V.~Kuzmin,
``The Hamiltonian formulation of N-bein, Einstein-Cartan, gravity in any dimension: The Progress Report,''
[arXiv:0907.1553 [gr-qc]].

\bibitem{Shapiro:2014kma}
I.~L.~Shapiro and P.~M.~Teixeira,
``Quantum Einstein-Cartan theory with the Holst term,''
Class. Quant. Grav. \textbf{31} (2014), 185002
doi:10.1088/0264-9381/31/18/185002
[arXiv:1402.4854 [hep-th]].

\bibitem{Plebanski:1977zz}
J.~F.~Plebanski,
``On the separation of Einsteinian substructures,''
J. Math. Phys. \textbf{18} (1977), 2511-2520
doi:10.1063/1.523215

\bibitem{Capovilla:1991qb}
R.~Capovilla, T.~Jacobson, J.~Dell and L.~J.~Mason,
``Selfdual two forms and gravity,''
Class. Quant. Grav. \textbf{8} (1991), 41-57
doi:10.1088/0264-9381/8/1/009

\bibitem{Krasnov:2016emc}
K.~Krasnov,
``Self-Dual Gravity,''
Class. Quant. Grav. \textbf{34} (2017) no.9, 095001
doi:10.1088/1361-6382/aa65e5
[arXiv:1610.01457 [hep-th]].

\bibitem{Bern:1998xc}
Z.~Bern, L.~J.~Dixon, M.~Perelstein and J.~S.~Rozowsky,
``One loop n point helicity amplitudes in (selfdual) gravity,''
Phys. Lett. B \textbf{444} (1998), 273-283
doi:10.1016/S0370-2693(98)01397-5
[arXiv:hep-th/9809160 [hep-th]].

\bibitem{Krasnov:2021cva}
K.~Krasnov and E.~Skvortsov,
``Flat Self-dual Gravity,''
doi:10.1007/JHEP08(2021)082
[arXiv:2106.01397 [hep-th]].

\bibitem{Abou-Zeid:2005zfo}
M.~Abou-Zeid and C.~M.~Hull,
``A Chiral perturbation expansion for gravity,''
JHEP \textbf{02} (2006), 057
doi:10.1088/1126-6708/2006/02/057
[arXiv:hep-th/0511189 [hep-th]].


\bibitem{Krasnov:2020bqr}
K.~Krasnov and Y.~Shtanov,
``Chiral perturbation theory for GR,''
JHEP \textbf{09} (2020), 017
doi:10.1007
[arXiv:2007.00995 [hep-th]].

\bibitem{Groh:2013oaa}
K.~Groh, K.~Krasnov and C.~F.~Steinwachs,
``Pure connection gravity at one loop: Instanton background,''
JHEP \textbf{07} (2013), 187
doi:10.1007
[arXiv:1304.6946 [hep-th]].

\bibitem{Krasnov:2015kva}
K.~Krasnov,
``One-loop beta-function for an infinite-parameter family of gauge theories,''
JHEP \textbf{03} (2015), 030
doi:10.1007
[arXiv:1501.00849 [hep-th]].

\bibitem{Christensen:1979iy}
S.~M.~Christensen and M.~J.~Duff,
``Quantizing Gravity with a Cosmological Constant,''
Nucl. Phys. B \textbf{170} (1980), 480-506
doi:10.1016/0550-3213(80)90423-X

\bibitem{Camporesi:1995fb}
R.~Camporesi and A.~Higuchi,
``On the Eigen functions of the Dirac operator on spheres and real hyperbolic spaces,''
J. Geom. Phys. \textbf{20} (1996), 1-18
doi:10.1016/0393-0440(95)00042-9
[arXiv:gr-qc/9505009 [gr-qc]].

\bibitem{Rigouzzo:2023sbb}
C.~Rigouzzo and S.~Zell,
``Coupling Metric-Affine Gravity to the Standard Model and Dark Matter Fermions,''
[arXiv:2306.13134 [gr-qc]].

\bibitem{Vassilevich:2003xt}
D.~V.~Vassilevich,
``Heat kernel expansion: User's manual,''
Phys. Rept. \textbf{388} (2003), 279-360
doi:10.1016/j.physrep.2003.09.002
[arXiv:hep-th/0306138 [hep-th]].

\bibitem{Besse:1987pua}
A.~L.~Besse,
``Einstein Manifolds,''
Springer-Verlag, 1987,
ISBN 978-0-387-15279-0

\bibitem{Krasnov:2020lku}
K.~Krasnov,
``Formulations of General Relativity,''
doi:10.1017/9781108674652

\bibitem{Penrose}
Penrose R, Rindler W.,
"1984 Spinors and spacetime vol 1, UK: Cambridge
University Press."

\end{thebibliography}
\end{document}